\documentclass[aps,pra,showpacs,superscriptaddress]{revtex4}

\usepackage{graphicx}

\begin{document}

\title{Quench Induced Growth of Distant Entanglement from Product and Locally Entangled States in Spin Chains}

\author{Bedoor Alkurtass}
\affiliation{Department of Physics and Astronomy, University College London, Gower Street, WC1E 6BT London, United Kingdom}
\affiliation{Department of Physics and Astronomy, King Saud University, Riyadh 11451, Saudi Arabia}
\author{Hannu Wichterich}
\affiliation{Department of Physics and Astronomy, University College London, Gower Street, WC1E 6BT London, United Kingdom}
\author{Sougato Bose}
\affiliation{Department of Physics and Astronomy, University College London, Gower Street, WC1E 6BT London, United Kingdom}

\begin{abstract}
We study the problem of entangling two spins at the distant ends of a spin chain by exploiting the nonequilibrium dynamics of the system after a sudden global quench. As initial states we consider a canted/spiral order product state of the spins, and singlets of neighboring pairs of spins. We find that within the class of canted order initial states, {\em no} entanglement is generated at any time {\em except} for the special case of the N\'eel state. While an earlier work had shown that the N\'eel state is indeed an excellent starting resource for the dynamical generation of long distance entanglement, the curious fact that this is the sole point within a large class of initial product states of the spins was not noted. On the other hand, we find that an initial state which is a series of nearest neighbor Bell states, and well motivated by some physical realizations, is also a good starting resource for end to end entanglement in a similar way to the N\'eel state. The scheme is shown to be robust to random single spin flip in the initial N\'eel state as well as randomness of the couplings.
\end{abstract}
\maketitle

Entanglement induced between distinct parts of a many-body system due to the nonequilibrium dynamics following quenches have attracted much attention recently. This has been largely motivated by experimental developments which allow for the realization of long time nonequilibrium dynamics -- notably with ultracold atoms in optical lattices \citep{kuhr,Trotzky,impurity,bound}. The potential for realizing such dynamics in ion traps has been also pointed out in both theoretical studies \citep{Hauke,daley,duan}, as well as the realization of spin chains \citep{christof1,richerme}. Some studies have examined the time development of entanglement between complementary blocks of a one dimensional spin system \citep{daley,Calabrese, deChiara, Das, Barmettler} as well as between {\em proximal} spins \citep{Sengupta,Alkurtass}. However, in the field of quantum information, it is most desirable to achieve long range entanglement, i.e., the entanglement between two individual spins separated by a significant distance. For example, it is essential for teleportation of a quantum state over a distance \citep{Bennett} and thereby to link well separated quantum registers. Motivated by this point, studies have also been made of the creation of long-range entanglement between the end spins of a spin chain through the nonequilibrium dynamics that follows a quench. The first study in this context considered a chain of harmonic oscillators which were suddenly coupled \citep{Eisert}. For spin chains, the first study in this context considered a sudden quench in the Ising anisotropy parameter of a $XXZ$ model \citep{Wichterich}. This was a {\em global quench} in the sense that all couplings of the Hamiltonian of a system were changed in the {\em same way} by the quench. Such quenches may be easier to realize in systems with limited local addressability such as for ultracold atoms in optical lattices. However, one can also suddenly change a local coupling, as may be easier in gate controlled systems such as quantum dots coupled to leads, and for a spin chain version of such systems, called Kondo systems, the mechanism to generate entanglement has also been worked out \citep{Sodano}. Other mechanisms to generate entanglement between distant spins may involve kicked systems \citep{Boness,Zueco} or long distance state swap induced quantum gates \citep{Banchi,Yao}, which also seems feasible in optical lattices and solid state spin chains in diamonds (see also alternative mechanism for entanglement distribution through the latter in Ref.\citep{yuting}). Alternatively, there are also various {\em static mechanisms} for generating long distance entanglement between the end spins of a spin chain \citep{Campos-Venuti}-\citep{statics-dyn}. Here it is also worth noting that generating entanglement between the end spins of a spin chain is a slightly different problem than directly using a spin chain as a data-bus for transferring quantum states, on which much work has been done (e.g. \citep{Bose}-\citep{Giovannetti}). Without really going into the details of the comparison of the various methods for the generation of long distance entanglement, here we wish to consider the question of whether entanglement can be generated between distant spins while starting off with a {\em product} state or a state were entanglement is initially very local (e.g. a product of Bell states). The mechanism will be the unitary dynamics induced by a global quench. Within a wide class of product states, we will find that a certain state emerges as the {\em only one} which generates a non-zero entanglement between distant spins. This study illustrates how important the initial magnetic order in the spin chain is, if one is to generate a long distance entanglement from a global quench. While such sensitivity is not good from a practical point of view, it is nonetheless really important to be aware of this so that one is adequately careful about the preparation of the initial state in any physical realization of this protocol.  On the other hand we will find that starting with a linear array of local singlets is also a good starting point for the generation of long distance entanglement and this might for certain realizations, provide a more practical starting point.\\

 An earlier work \citep{Wichterich} considered a one dimensional chain of $N$ spin-$1/2$ systems described by the $XXZ$ Hamiltonian,
\begin{equation}
H_{\text{XXZ}}=\frac{J}{2} \sum_{k=1}^{N-1} \left(\hat{\sigma}_{k}^{x} \hat{\sigma}_{k+1}^{x}+\hat{\sigma}_{k}^{y} \hat{\sigma}_{k+1}^{y} +\Delta \hat{\sigma}_{k}^{z} \hat{\sigma}_{k+1}^{z} \right) \, ,
\label{Hamiltonian}
\end{equation}
where $J$ is the coupling strength, $\Delta$ is the anisotropy parameter and $\sigma^{x}_k, \sigma^{y}_k$ and $\sigma^{z}_k$ are the Pauli operators acting on a site $k$. It was shown that quenching the anisotropy parameter from $\Delta\rightarrow\infty$, which corresponds to a N\'eel state, to $\Delta=0$, which is the $XX$ Hamiltonian, resulted in the evolution of the end spins to a highly entangled state without requiring any local control of the spins. It is precisely to generalize this work that we consider two initial states: a canted/spiral order product state and a product of singlets. The preparation of the initial states as the ground state of a Hamiltonian is now not our principal concern, but it is clearly possible with uncoupled and dimerized Hamiltonians. With the system placed in the above initial states by some means, we assume that a $XX$ Hamiltonian is switched on at time $t=0$. The system then evolves according to the $XX$ Hamiltonian. We then study analytically the entanglement generated between the distant ends of the chain due to the dynamics following our quench (here by quench we mean the sudden switching on of the couplings so as to make the system behave as a $XX$ spin chain).

\section{Canted Order Initial state}
In Ref.\citep{Wichterich}, the N\'eel state was considered because it is the natural ground state for $\Delta \rightarrow \infty$. However, here we are interested in the theoretical (fundamental) problem of how much entanglement between distant spins can be generated from separable states and a global switch of couplings. The whole class of separable states is enormous as well as the class of product states of several spins. Thus we examine here the canted order pure initial state
\begin{equation}
\left|\psi\right\rangle=\mathop{\otimes}_{k=1}^{N} \left|\psi_k\right\rangle=\mathop{\otimes}_{k=1}^{N} \left( \cos{\frac{\theta_{k}}{2}} \left|\uparrow_{k}\right\rangle + \sin{\frac{\theta_{k}}{2}} \left|\downarrow_{k}\right\rangle \right) \, ,
\label{CantedState}
\end{equation}
where $\theta_k=(k-1) \alpha$ and $0\leq\alpha\leq 2\pi$. This initial state may be prepared in several ways. The most obvious way is to have an {\em uncoupled} Hamiltonian and a magnetic field varying in direction from site to site which tilts spins accordingly. Even with a magnetic field varying only in one (say the $x$) direction, one can prepare the above state for $\alpha \leq \frac{\pi}{2(N-1)}$ as the ground state of the coupled Hamiltonian
\begin{equation}
H_{\text{canted}}=-\sum_{k=1}^{N-1} \sigma_k^z \otimes \sigma_{k+1}^z + \sum_{k=1}^N h_k^x \sigma_k^x -h_1^z \sigma_1^z \, .
\end{equation}
With $h_k^x$ increasing with the site number, the ground state is a chain in which each spin is tilted counter-clockwise with an angle $\alpha$ compared with the left neighboring spin (Fig.~\ref{CantedState}). After N\'eel order, a canted order is a next level of generalization that we can make for an initial product state of a spin chain. 

\section{The Quench and the Generation of Entanglement}
\label{canted}
 We are motivated by the scope of solving for the entanglement of the distant end spins analytically for the given initial state.
It is with this view of making analytic progress that we let the system with the canted order initial state evolve with the $XX$ Hamiltonian. One can regard this as an instantaneous change (or a quench) from an uncoupled Hamiltonian with local magnetic fields or the Hamiltonian $H_{\text{canted}}$ to the $XX$ Hamiltonian. Also note that in Ref.\citep{Wichterich} it was found that $H_{\text{XXZ}}$ with $\Delta \not= 0$ is worse in terms of generating entanglement in comparison to the case of $\Delta=0$. This can be regarded as another motivation for evolving the system with $XX$ Hamiltonian. So one first prepares a canted order initial state of a few spins by one of the means described in the previous section and then, at time $t=0$ suddenly switches on a $XX$ Hamiltonian interaction between them. We are interested in the entanglement produced between spins $1$ and $N$ as a function of time due to this $XX$ Hamiltonian, $H$. 
\begin{figure}[htbp]
 \centering 
\includegraphics[width=7cm]{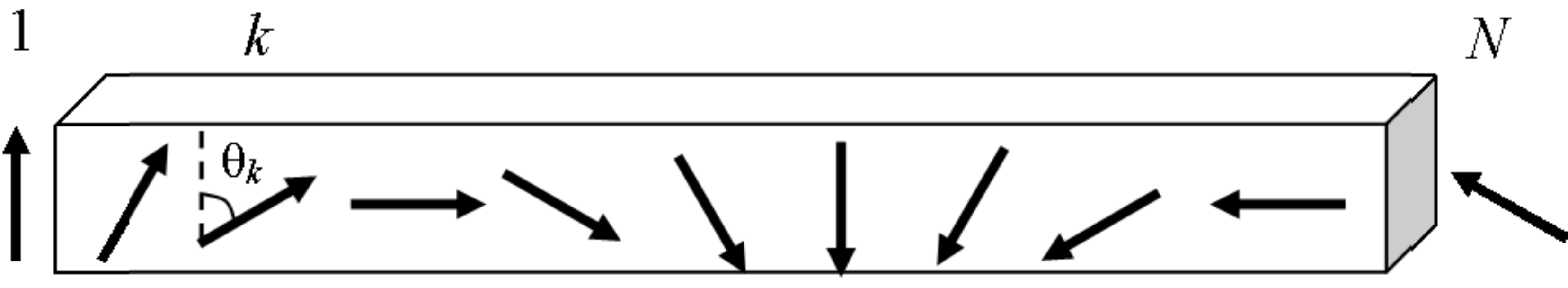}
   \caption{{\protect\footnotesize This figure shows the canted order state considered as an initial state (Eq.~\ref{CantedState}).}}
\end{figure}\\
To make progress, we first use the Jordan-Wigner transformation,
\begin{equation}
\hat{c}_{l}^{\dagger}=\hat{S}_{1,l-1} \hat{\sigma}_l^{+}, \quad \hat{c}_{l}= \hat{S}_{1,l-1} \hat{\sigma}_l^{-}
\end{equation}
with $\sigma^{\pm}=\frac{1}{2}(\sigma^{x}\pm\sigma^{y})$ and $\hat{S}_{l,m}=\mathop{\otimes}_{k=l}^{m} (- \hat{\sigma}_k^{z})$,
to write the Hamiltonian in terms of fermionic operators and then diagonalize $H$ to get the fermionic operators in the Heisenberg picture
\begin{equation}
\hat{c}_{k}(t)=\sum_{l} f_{k,l} \hat{c}_{l}(0), \quad \hat{c}_{k}^{\dagger}(t)=\sum_{l} f_{k,l}^{*} \hat{c}_{l}^{\dagger}(0) \, ,
\end{equation}
\begin{equation}
f_{k,l}(t)= \frac{2}{N+1}\sum_{m=1}^{N} \sin(q_m k) \sin(q_m l) e^{-i \epsilon_m t} \, ,
\end{equation}
with $\epsilon_m=2 J \cos(q_m)$ and $q_m=\frac{\pi m}{N+1}$.
To study the entanglement generated between the distant ends of the chain ($1, N$) we find the reduced density matrix in the basis $\left\{\left| \uparrow\uparrow\right\rangle, \left| \uparrow\downarrow\right\rangle,\left| \downarrow\uparrow\right\rangle,\left| \downarrow\downarrow\right\rangle\right\}$
\begin{equation}
\hat{\rho}_{1,N}(t)=\left(\begin{array}{cccc}
\left\langle \hat{P}_{1}^{\uparrow} \hat{P}_{N}^{\uparrow} \right\rangle & \left\langle \hat{P}_{1}^{\uparrow} \hat{\sigma}_{N}^{-} \right\rangle & \left\langle \hat{\sigma}_{1}^{-} \hat{P}_{N}^{\uparrow} \right\rangle & \left\langle \hat{\sigma}_{1}^{-} \hat{\sigma}_{N}^{-} \right\rangle\\
\left\langle \hat{P}_{1}^{\uparrow} \hat{\sigma}_{N}^{+} \right\rangle & \left\langle \hat{P}_{1}^{\uparrow} \hat{P}_{N}^{\downarrow} \right\rangle & \left\langle \hat{\sigma}_{1}^{-} \hat{\sigma}_{N}^{+} \right\rangle & \left\langle \hat{\sigma}_{1}^{-} \hat{P}_{N}^{\downarrow} \right\rangle\\
\left\langle \hat{\sigma}_{1}^{+} \hat{P}_{N}^{\uparrow} \right\rangle & \left\langle \hat{\sigma}_{1}^{+} \hat{\sigma}_{N}^{-} \right\rangle & \left\langle \hat{P}_{1}^{\downarrow} \hat{P}_{N}^{\uparrow} \right\rangle & \left\langle \hat{P}_{1}^{\downarrow} \hat{\sigma}_{N}^{-} \right\rangle\\
\left\langle \hat{\sigma}_{1}^{+} \hat{\sigma}_{N}^{+} \right\rangle & \left\langle \hat{\sigma}_{1}^{+} \hat{P}_{N}^{\downarrow} \right\rangle & \left\langle \hat{P}_{1}^{\downarrow} \hat{\sigma}_{N}^{+} \right\rangle & \left\langle \hat{P}_{1}^{\downarrow} \hat{P}_{N}^{\downarrow} \right\rangle
\end{array}\right) \, ,
\label{rho1N}
\end{equation}
where $\hat{P}_{l}^{\downarrow} = \hat{\sigma}_{l}^{-}\hat{\sigma}_{l}^{+},\hat{P}_{l}^{\uparrow} = \hat{\sigma}_{l}^{+}\hat{\sigma}_{l}^{-}$, $\left\langle ... \right\rangle=Tr(\rho(t)...)$ and $\rho(t)=e^{-i \hat{H} t} \rho(0) e^{i \hat{H} t}$.
We will show the evaluation of one matrix element, say $(1,4)$, and the rest of the elements are found similarly. The matrix element $(1,4)$ is
\begin{equation}
\left\langle \hat{\sigma}_{1}^{-} \hat{\sigma}_{N}^{-} \right\rangle= \left\langle \hat{c}_1 (t) \hat{c}_N (t) \hat{S}_{1,N-1}\right\rangle= -\left\langle \hat{c}_1(t) \hat{c}_N (t) \hat{S}_{1,N}\right\rangle \, ,
\end{equation}
where we used $\hat{\sigma}_{l}^{-} \hat{S}_{l}=-\hat{\sigma}_{l}^{-}$. Hence
\begin{equation}
\left\langle \hat{\sigma}_{1}^{-} \hat{\sigma}_{N}^{-} \right\rangle=-\sum_{l,m} f_{1,l}(t) f_{N,m}(t) \left\langle \hat{c}_l(0) \hat{c}_m (0) \hat{S}_{1,N} \right\rangle.
\end{equation}
Now we have 3 cases, first if $l=m$, then
\begin{equation}
\left\langle \hat{c}_l(0) \hat{c}_m (0) \hat{S}_{1,N} \right\rangle=0.
\end{equation}
Second, If $l>m$, then
\begin{eqnarray}
\nonumber \left\langle \hat{c}_l(0) \hat{c}_m (0) \hat{S}_{1,N} \right\rangle&=&\left\langle \hat{S}_{1,l-1} \hat{\sigma}_{l}^{-}(0) \hat{S}_{1,m-1} \hat{\sigma}_{m}^{-}(0) \hat{S}_{1,N}\right\rangle\quad\\
\nonumber &=&\left\langle \psi_l \right| \hat{\sigma}_{l}^{-}(0) \left| \psi_l\right\rangle \left\langle\psi_m \right|\hat{\sigma}_{m}^{-}(0) \left| \psi_m\right\rangle \mathop{\otimes}_{k=1}^{m-1}\left\langle\psi_k \right| -\sigma^{z}_{k}\left| \psi_k\right\rangle \mathop{\otimes}_{k=l+1}^{N}\left\langle\psi_k \right| -\sigma^{z}_{k}\left| \psi_k\right\rangle\quad\\
&=&\frac{1}{4} \sin \theta_l \sin \theta_m \mathop{\otimes}_{k=1}^{m-1}(-\cos \theta_k) \mathop{\otimes}_{k=l+1}^{N}(-\cos \theta_k).\quad
\end{eqnarray}

where we used $(\sigma^z_k)^2=1$. Finally, if $l<m$, then
\begin{eqnarray}
\nonumber \left\langle \hat{c}_l(0) \hat{c}_m (0) \hat{S}_{1,N} \right\rangle&=&\left\langle \hat{S}_{1,l-1} \hat{\sigma}_{l}^{-} (0) \hat{S}_{1,m-1} \hat{\sigma}_{m}^{-} (0)  \hat{S}_{1,N} \right\rangle\quad\\
\nonumber &=&\left\langle \psi_l \right| \hat{\sigma}_{l}^{-}(0) \left| \psi_l\right\rangle \left\langle\psi_m \right|\hat{\sigma}_{m}^{-}(0) \left| \psi_m\right\rangle\mathop{\otimes}_{k=1}^{l-1}\left\langle\psi_k \right| -\sigma^{z}_{k}\left| \psi_k\right\rangle\mathop{\otimes}_{k=m+1}^{N}\left\langle\psi_k \right| -\sigma^{z}_{k}\left| \psi_k\right\rangle\quad\\
 &=&\frac{1}{4} \sin \theta_l \sin \theta_m \mathop{\otimes}_{k=1}^{l-1}(-\cos \theta_k)\mathop{\otimes}_{k=m+1}^{N}(-\cos \theta_k).\quad
\label{analytic}
\end{eqnarray}
Similarly we can find analytic expressions for all the matrix elements of $\rho_{1,N}$. However, these expressions, of the form of Eq.(\ref{analytic}), have themselves to be evaluated numerically by virtue of the long products involved. This evaluation can be carried out efficiently even for very large values of $N$.

\begin{figure}[htbp]
\centering 
\includegraphics[width=14cm]{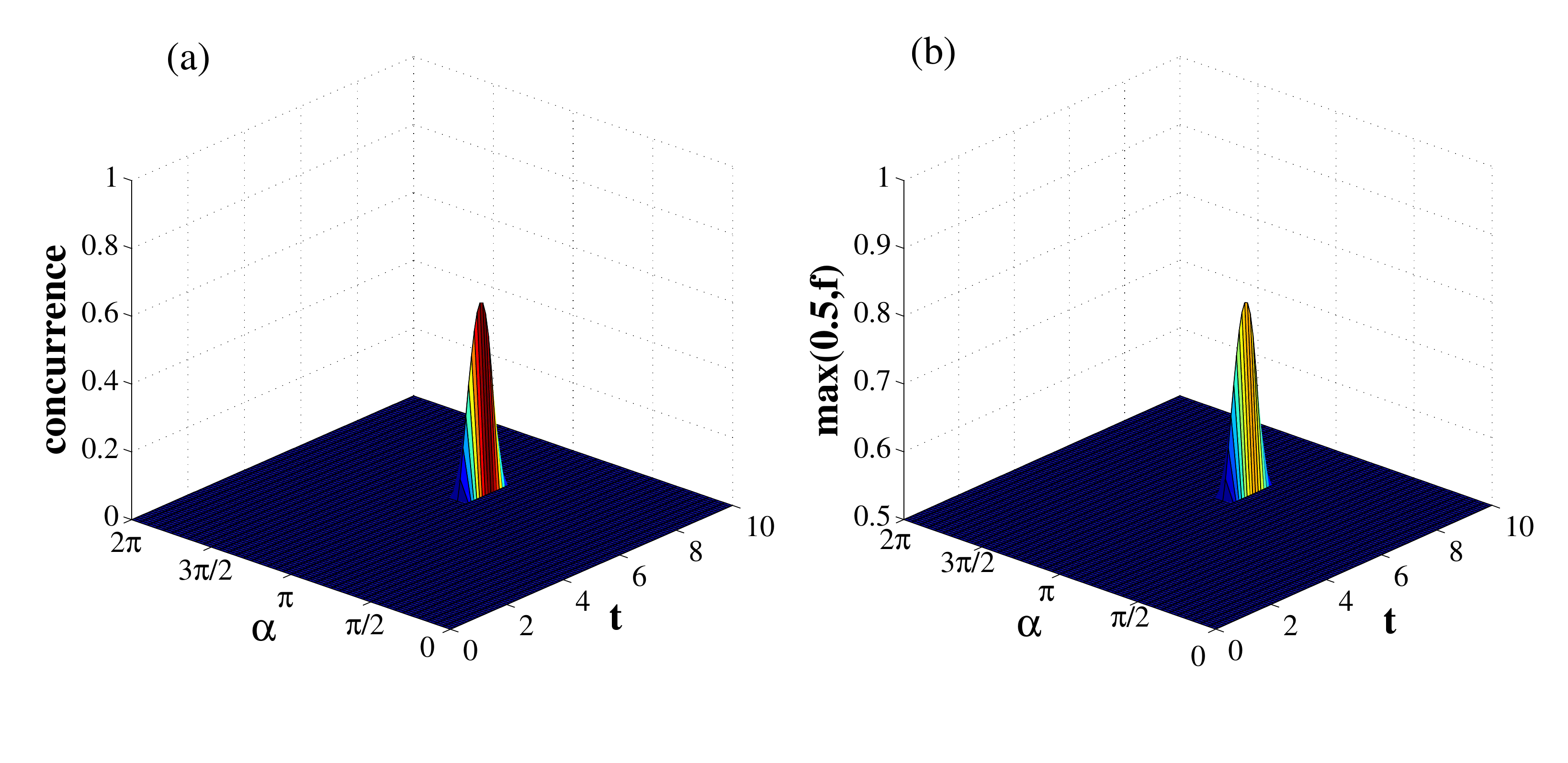}
   \caption{{\protect\footnotesize (Color online) Dynamics of the concurrence and the fully entangled fraction, of spins $1$ and $N$ of the chain for $N=24$ as a function of the canting angle $\alpha$. The scaled time $t$ is in units of $1/J$.}}
\label{CantedOrderN24}
\end{figure}

\begin{figure}[htbp]
\centering 
\includegraphics[width=14cm]{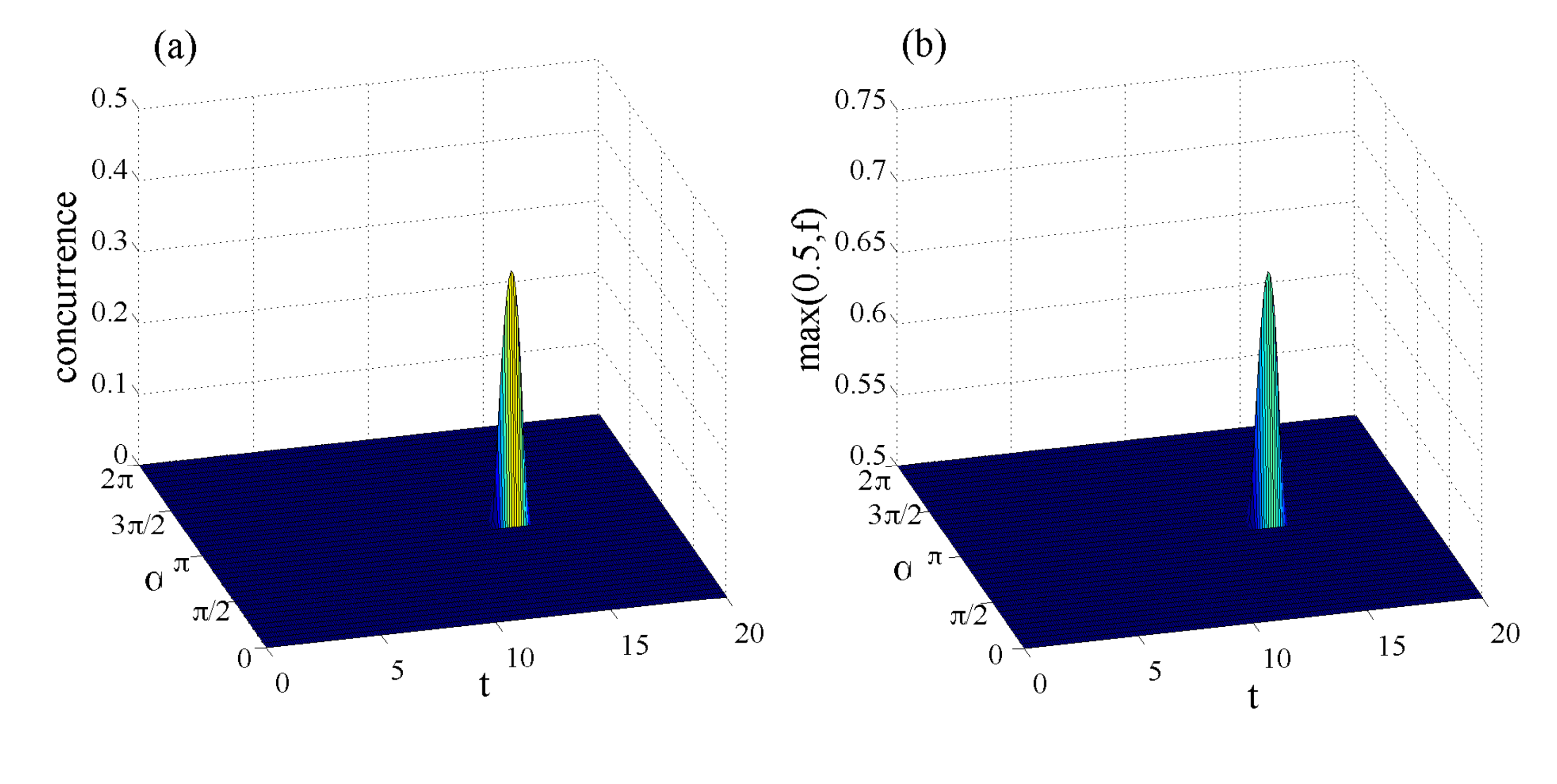}
   \caption{{\protect\footnotesize (Color online) Dynamics of the concurrence and the fully entangled fraction, of spins $1$ and $N$ of the chain for $N=50$ as a function of the canting angle $\alpha$. The scaled time $t$ is in units of $1/J$.}}
\label{CantedOrderN50}
\end{figure}

\begin{figure}[htbp]
\centering 
{\includegraphics[width=7cm]{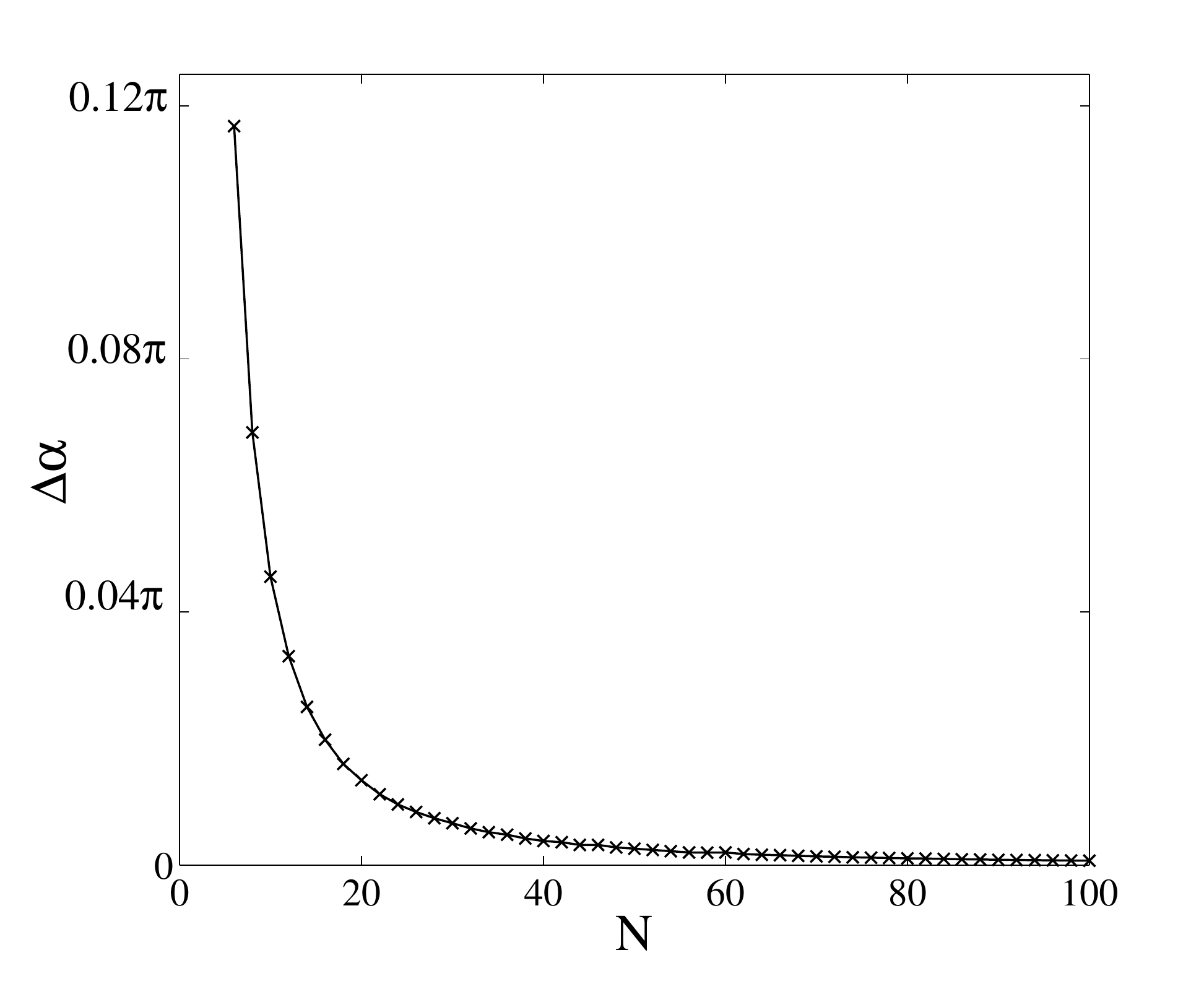}}\caption{{\protect\footnotesize (Color online) The full width at half maximum $\Delta \alpha$ of the peak in entanglement around $\alpha=\pi$ at the optimal time versus the chain length.}}
\label{PeakWidth}
\end{figure}

After having determined the elements of $\rho_{1,N}$, we are now in a position to compute the entanglement between spins $1$ and $N$ as a function of time. One way of quantifying the entanglement is to compute the concurrence, which is a measure of entanglement for two qubits \citep{wootters}. However, note that this entanglement will, in the end, be used for teleporting arbitrary quantum states between two registers so that it is the ``fully entangled fraction" rather than some measure of entanglement, which is the more relevant figure of merit. One would first apply the entanglement purification protocol in which a given number of copies of the state $\rho_{1,N}$ can be converted, by local actions at both ends and classical communication, to a smaller number of maximally entangled states which can then be used for teleportation \citep{Bennett2}. In order to test the possibility of the purification procedure we use the {\em fully entangled fraction} defined as
\begin{equation}
f=max(\left\langle e|\rho|e \right\rangle) \, ,
\end{equation}
where the maximum is taken over all maximally entangled states $\left\{|e \rangle\right\}$. The purification procedure is possible if $f>\frac{1}{2}$ \citep{Bennett2}. Moreover, if we were to use the state $\rho_{1,N}$ itself for teleportation, then the fidelity of the teleported state is given by $F=\frac{2f+1}{3}$ \citep{Horodecki}. We will thus also compute the fully entangled fraction $f$ for $\rho_{1,N}$ as a function of time and the canting angle $\alpha$, as this is much more indicative of the eventual application in connecting quantum registers, namely teleportation. \\
Figs~\ref{CantedOrderN24} and \ref{CantedOrderN50} show the dynamics of the fully entangled fraction and the concurrence of $\rho_{1,N}$ as a function of $\alpha$ for $N=24$ and $N=50$ spins respectively. Clearly, at around a scaled time very close to $t_{\text{opt}}\sim \frac{N}{4J}$ (in fact the time is slightly larger than $t_{\text{opt}}$) and for the N\'eel state, a state $\rho_{1,N}$ with the highest fully entangled fractions $f\sim 0.78$ (for $N=24$) and $f\sim 0.68$ (for $N=50$) is generated, which will not only directly allow one to teleport a state with a fidelities of $F\sim 0.85$ (for $N=24$) and  $F\sim 0.79$ (for $N=50$), but also enable entanglement purification (using many copies of $\rho_{1,N}$) to obtain a state giving perfect teleportation fidelity. We focused here on the first emergence in time of long range entanglement. We note the sharp peak in the entanglement at $\alpha=\pi$ which corresponds to the N\'eel state. This means that the proposed scheme generates entanglement for a N\'eel state and not for the general canted order state. The narrowness of the peak in $\alpha$, Fig.~\ref{PeakWidth}, is especially striking implying essentially that unless a canted order state is {\em extremely close} to the N\'eel state it is ill-suited for generating any long distance entanglement. This uniqueness of the N\'eel state among all the canted order states is one of our principal findings that adds to the knowledge with respect to Ref.\citep{Wichterich}, where the high entanglement generating ability of only the N\'eel state was noted. We will attempt to provide an explanation of this effect in the section on discussions in a later part of the paper.

\section{Series of Bell States as an Initial State}
\label{BS}
There may be other states which are easy to prepare as initial states in practice in various physical realizations. As long as their entanglement is not long range to start with, the quench induced dynamics still serves its purpose of generating long distance entanglement. We can thus look at an initial state which is a product of states where spins are entangled only with their nearest neighbors, namely a product of Bell states. We now consider the following product of Bell states as the initial state
\begin{equation}
\left|\psi\right\rangle=\mathop{\otimes}_{k=1}^{N/2} \left|\psi_{k,k+1}\right\rangle = \mathop{\otimes}_{k=1}^{N/2} \left(\left|\uparrow_{k} \downarrow_{k+1}\right\rangle - \left|\downarrow_{k} \uparrow_{k+1}\right\rangle \right).
\end{equation}
The above state can be implemented using cold atoms in an optical superlattice formed by two independent lattices with different periods. Applying a spin-dependent energy offset results in pairs of singlets. \citep{Yang,Rey}\\
Starting with the above initial state, the system then evolves according to the $XX$ Hamiltonian (Eq.(\ref{Hamiltonian}) with $\Delta=0$). Using Eq.(\ref{rho1N}) and
\begin{equation}
\left\langle \psi_{l,l+1} \right|\hat{\sigma}_{l}^{\pm}\left|\psi_{l,l+1}\right\rangle=0, \left\langle \psi_{l,l+1} \right|\hat{\sigma}_{l+1}^{\pm}\left|\psi_{l,l+1}\right\rangle=0, \left\langle\hat{\sigma}_{1}^{\pm}\hat{\sigma}_{N}^{\pm} \right\rangle=0,
\end{equation}
we find that the only non-vanishing elements of the reduced density matrix are
\begin{equation}
\hat{\rho}_{1,N}(t)=\left(\begin{array}{cccc}
a(t) & & & \\
 & b(t) & c(t) & \\
 & c'(t) & b'(t) & \\
& & & a'(t)
\end{array}\right) \, .
\end{equation}
The matrix elements are
\begin{eqnarray}
a(t)&=&\left\langle \hat{c}_{N}^{\dagger}(t)\hat{c}_{N}(t) \right\rangle \left\langle \hat{c}_{1}^{\dagger}(t)\hat{c}_{1}(t) \right\rangle - \left\langle \hat{c}_{N}^{\dagger}(t)\hat{c}_{1}(t) \right\rangle \left\langle \hat{c}_{1}^{\dagger}(t)\hat{c}_{N}(t) \right\rangle\\
a'(t)&=&1 - \left\langle \hat{c}_{1}^{\dagger}(t)\hat{c}_{1}(t) \right\rangle- \left\langle \hat{c}_{N}^{\dagger}(t)\hat{c}_{N}(t) \right\rangle + a (t)\\
b(t)&=&\left\langle \hat{c}_{1}^{\dagger}(t)\hat{c}_{1}(t)\right\rangle - a(t)\\
b'(t)&=&\left\langle \hat{c}_{N}^{\dagger}(t)\hat{c}_{N}(t)\right\rangle - a(t)\\
c(t)&=&\sum_{l,m} f_{1,l}(t) f_{N,m}^{*} (t) \left\langle \hat{c}_l (0)\hat{c}_{m}^{\dagger} (0) S_{1,N}\right\rangle\\
c'(t)&=&\sum_{l,m} f_{N,l}(t) f_{1,m}^{*} (t) \left\langle \hat{c}_l (0)\hat{c}_{m}^{\dagger} (0) S_{1,N}\right\rangle
\end{eqnarray}
For the two point correlation function $\left\langle \hat{c}_{i}^{\dagger}(t)\hat{c}_{j}(t)\right\rangle=\sum_{l,m} f_{i,l}^{*}(t) f_{j,m}(t) \left\langle \hat{c}_{l}^{\dagger}(0)\hat{c}_{m}(0)\right\rangle$ we have 3 cases, first if $|{l-m}|>1$, then\begin{equation} \left\langle \hat{c}_{l}^{\dagger}(0) \hat{c}_m (0) \right\rangle=0.\end{equation}
Second, if $l=m$, then
\begin{equation}
\left\langle \hat{c}_{l}^{\dagger}(0) \hat{c}_l (0) \right\rangle=\left\langle \hat{\sigma}_{l}^{+}(0) \hat{\sigma}_{l}^{-}(0)\right\rangle=\frac{1}{2}
\end{equation}
Finally, if $|{l-m}|=1$, then
\begin{equation}
\left\langle \hat{c}_{l}^{\dagger} (0) \hat{c}_m (0) \right\rangle=\left\langle \hat{\sigma}_{l}^{+} (0)  \hat{\sigma}_{m}^{-} (0) \right\rangle=-\frac{1}{2}
\end{equation}
While for the term $\left\langle \hat{c}_l (0)\hat{c}_{m}^{\dagger} (0) S_{1,N}\right\rangle$ we have 4 cases, first if $l=m$, then
\begin{equation}
\left\langle \hat{c}_{l}(0) \hat{c}_{m}^{\dagger} (0) \hat{S}_{1,N} \right\rangle=(-1)^{N/2-1} \left\langle \hat{\sigma}_{l}^{-} \hat{\sigma}_{l}^{+} \hat{S}_{l+1}\right\rangle=\frac{1}{2} (-1)^{N/2}.
\end{equation}
Second, if $l=odd$ and $m=l+1$, then
\begin{equation}
\left\langle \hat{c}_{l}(0) \hat{c}_{m}^{\dagger} (0) \hat{S}_{1,N} \right\rangle=\left\langle \hat{\sigma}_{l}^{-} \hat{\sigma}_{l+1}^{+} \hat{S}_{1,l-1} \hat{S}_{l+2,N} \right\rangle=\frac{1}{2} (-1)^{N/2}.
\end{equation}
Third, if $l=even$ and $m=l-1$, then
\begin{equation}
\left\langle \hat{c}_{l}(0) \hat{c}_{m}^{\dagger} (0) \hat{S}_{1,N} \right\rangle=\left\langle \hat{\sigma}_{l}^{-} \hat{\sigma}_{l-1}^{+} \hat{S}_{1,l-2} \hat{S}_{l+1,N} \right\rangle=\frac{1}{2} (-1)^{N/2}.
\end{equation}
Finally, for all other cases
\begin{equation}
\left\langle \hat{c}_{l}(0) \hat{c}_{m}^{\dagger} (0) \hat{S}_{1,N} \right\rangle=0.
\end{equation}

\begin{figure}[htbp]
 \centering 
\includegraphics[width=14cm]{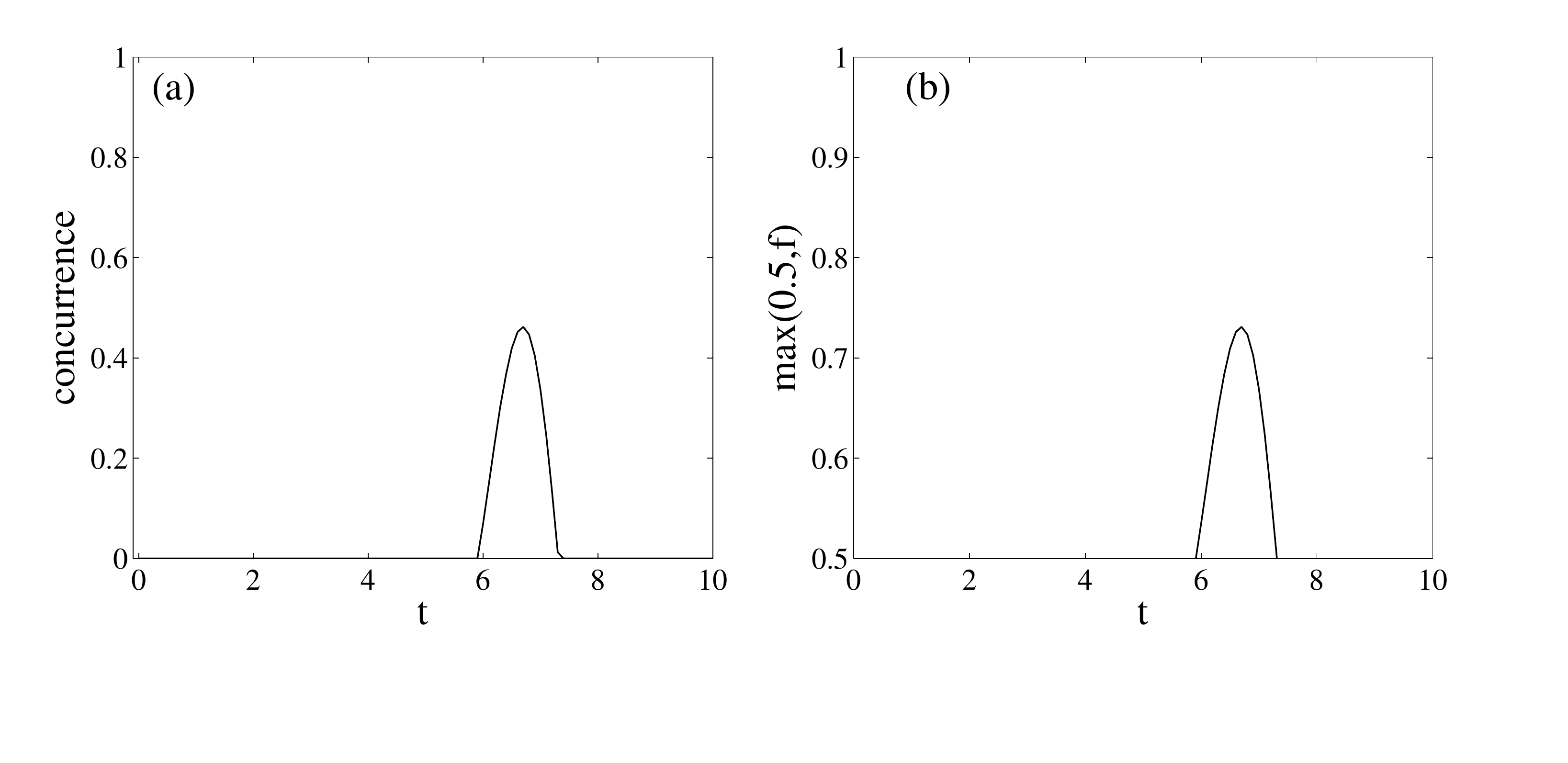}
   \caption{{\protect\footnotesize (Color online) Dynamics of the concurrence and the fully entangled fraction, of spins $1$ and $N$ of the chain for $N=24$ spins with a series of Bell pairs as an initial state. The scaled time $t$ is in units of $1/J$.}}
\label{Singlets}
\end{figure}

\begin{figure}[htbp]
 \centering 
\includegraphics[width=14cm]{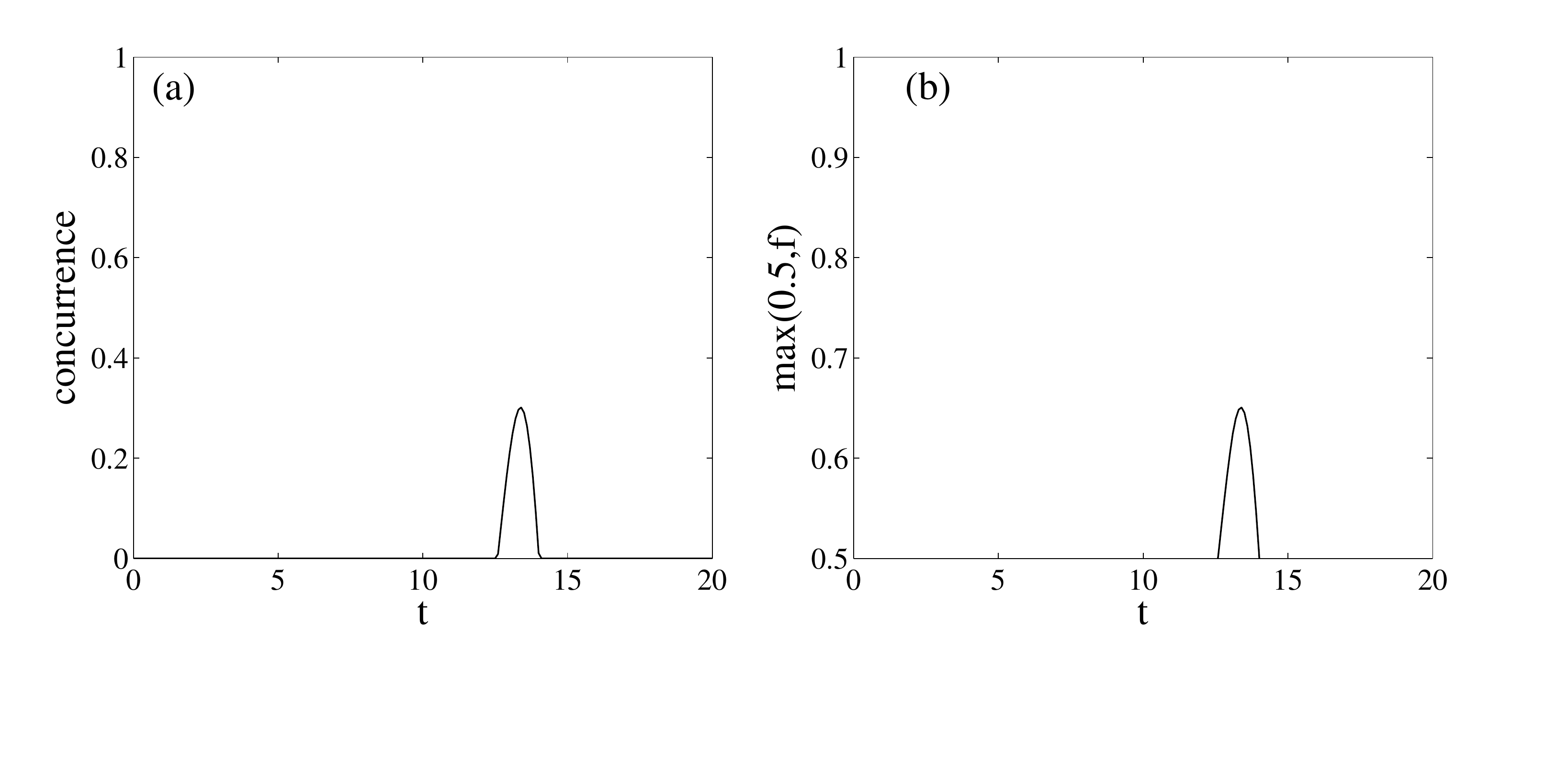}
   \caption{{\protect\footnotesize (Color online) Dynamics of the concurrence and the fully entangled fraction, of spins $1$ and $N$ of the chain for $N=50$ spins with a series of Bell pairs as an initial state. The scaled time $t$ is in units of $1/J$.}}
\label{Singlets50}
\end{figure}

Figs.~\ref{Singlets} and \ref{Singlets50} show the dynamics of the concurrence and the fully entangled fraction of the distant end spins of the chain for $N=24$ spins, i.e. $12$ Bell pairs and $N=50$ spins, i.e., $25$ Bell pairs respectively. We note that the behavior is similar to the case of N\'eel state. It shows that the initial state of a series of Bell pairs act very similarly to the N\'eel state in terms of entanglement generation (maximally entangled fractions of $0.73$ and $0.65$ at times close to (slightly higher than) $t_{\text{opt}}\sim \frac{N}{4J}$.

\begin{figure}[htbp]
 \centering 
\includegraphics[width=14cm]{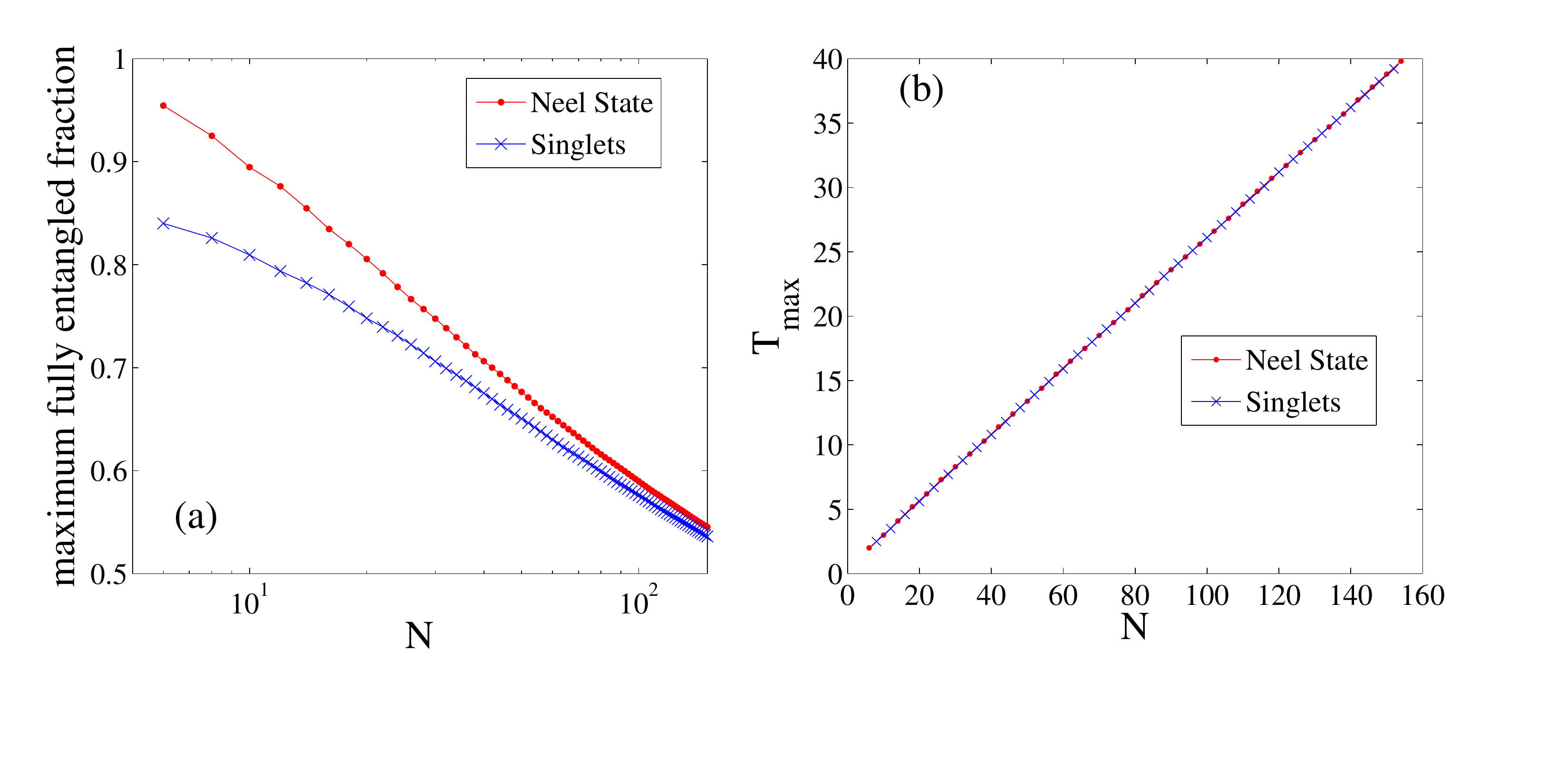}
   \caption{{\protect\footnotesize (Color online) (a) maximum fully entangled fraction of $\rho_{1,N}$ versus the chain length, (b) the time at which the first peak of entanglement occurs versus the chain length.}}
\label{Scaling}
\end{figure}

\begin{figure}
 \centering 
\includegraphics[width=12cm]{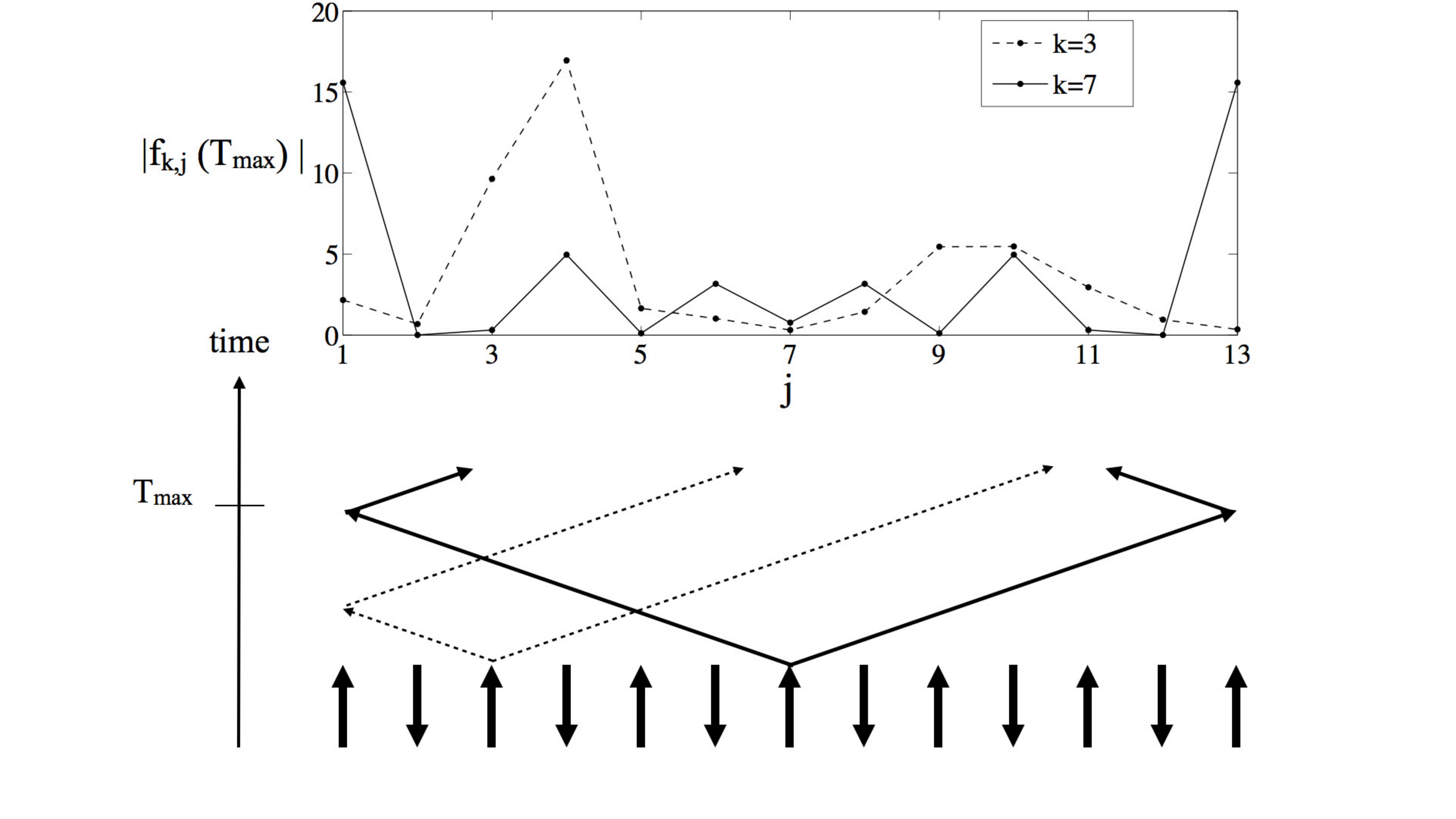}
   \caption{{\protect\footnotesize This figure shows the quantum walk of excitations (fermions) starting from an initial state (Eq.~\ref{CantedState}). The $k=7$ fermion is shown to generate an equal distribution of probabilities at either end at time $t\sim \frac{N}{4J}$, whereas the $k=3$ fermion generates a state with unequal probabilities at the ends.}}
 \label{expl}
\end{figure}

\section{Errors}
\begin{figure}
 \centering 
\includegraphics[width=12cm]{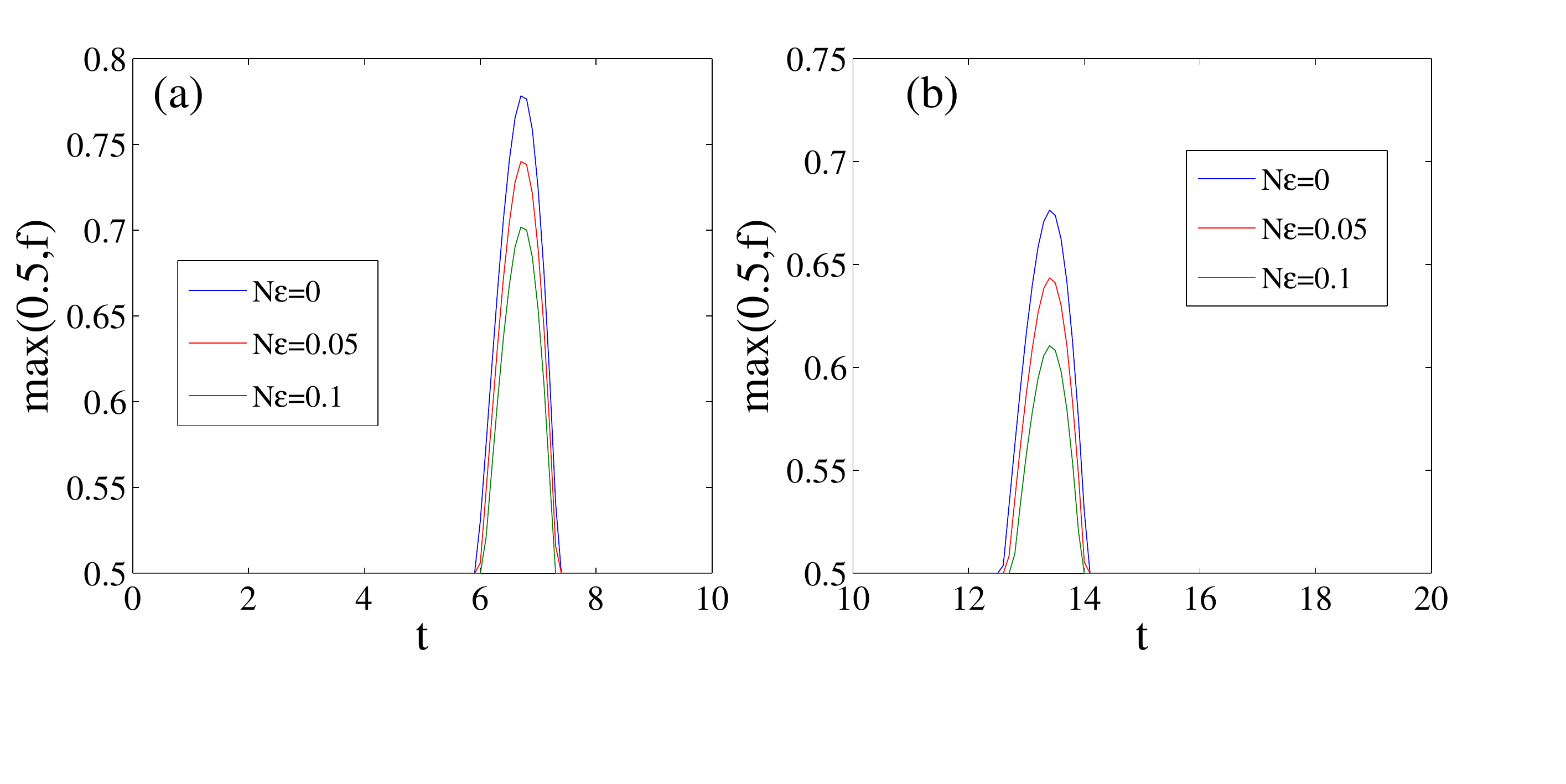}
   \caption{{\protect\footnotesize (color online) The fully entangled fraction for an initial N\'eel state with a random single spin flip with probability $N\epsilon=0,0.05,0.1,0.15$ (top to bottom) for (a) $N=24$, (b) $N=50$.}}
 \label{CantedDisorder}
\end{figure}

\begin{figure}
 \centering 
\includegraphics[width=12cm]{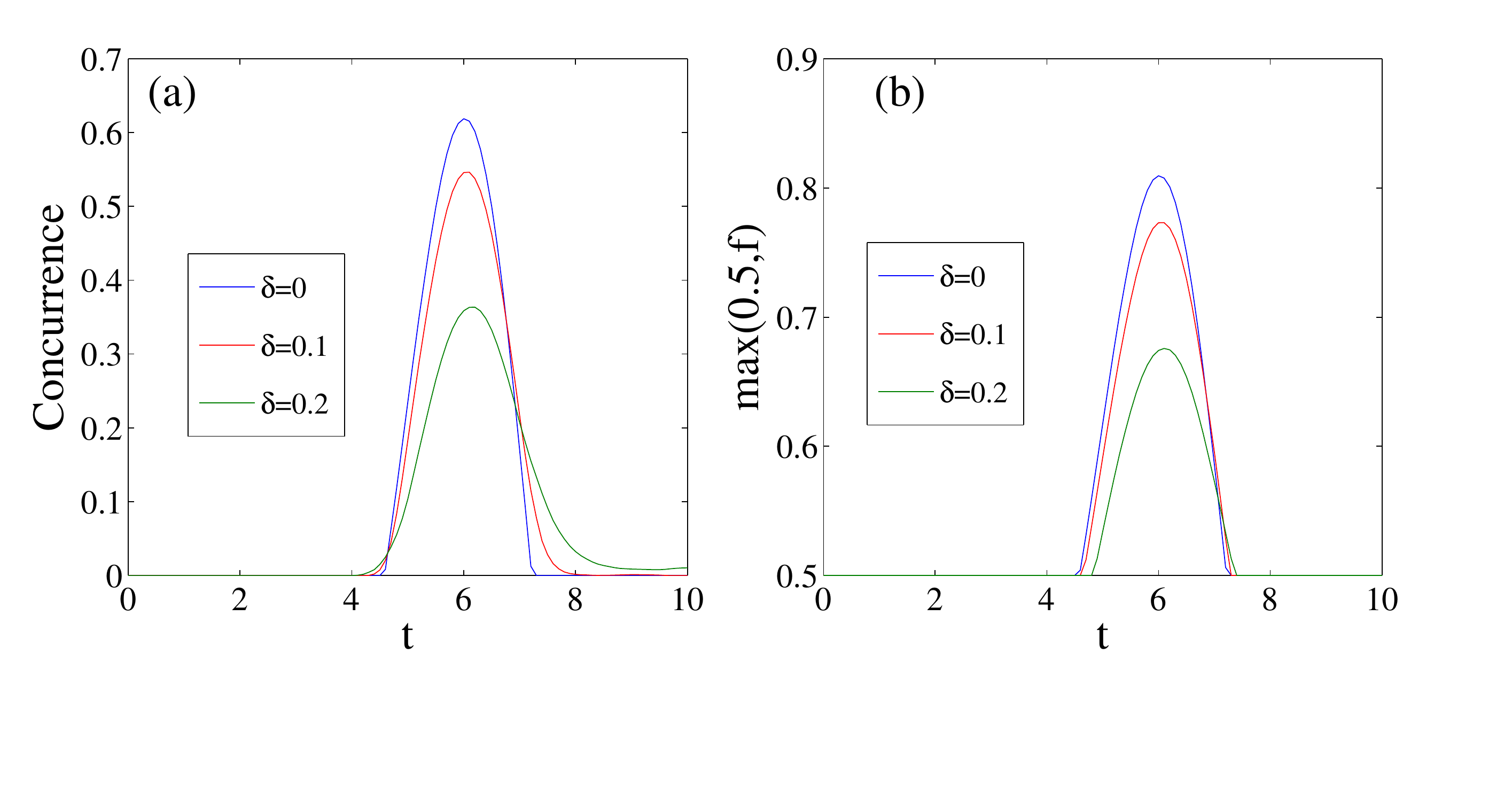}
   \caption{{\protect\footnotesize (color online) The concurrence and fully entangled fraction for a series of Bell pairs as an initial state with random offsets in the couplings through the chain $\delta=0,0.1,0.2$ (top to bottom) for $N=10$ (average taken over $100$ realizations).}}
 \label{SingletsDisorder}
\end{figure}

In any experimental implementation, some disorders will be present and might affect the resulting entanglement \cite{Tsomokos}. To measure the robustness of our scheme we consider two types of disorder. First we study the effect of a small disorder in the preparation of the initial N\'eel state. We consider a random single spin flip with a probability $N\epsilon$, i.e. the initial density matrix is
\begin{equation}
\rho=\left(1-N\epsilon\right)\left|\textrm{N\'eel}\right\rangle\left\langle \textrm{N\'eel}\right| + \epsilon \sum_{k=1}^{N} \sigma_k^x\left|\textrm{N\'eel}\right\rangle\left\langle \textrm{N\'eel}\right| \sigma_k^x
\end{equation}
Figure~\ref{CantedDisorder} shows the resulting fully entangled fraction for $N=24$ and $N=50$ with probabilities $N\epsilon=0,0.05,0.1,0.15$ (top to bottom). The entanglement at $T_{max}$ is not seriously affected with such disorder with a spin flip probability as large as $0.05$ and $T_{max}$ is also unchanged. In \cite{Wichterich} it was shown that this scheme for generating entanglement is also robust to randomness in the couplings throughout the chain for an initial N\'eel state. We consider the effect of such disorder for an initial series of Bell pairs. The coupling between neighboring sites $J$ is taken to be $J_k=J(1+\delta_k)$ with a normally distributed set $\delta_k$ with zero mean value and standard deviation $\delta$. Figure~\ref{SingletsDisorder} shows the resulting entanglement for $N=10$ and $\delta=0,0.1,0.2$ (top to bottom). The entanglement is not seriously suppressed for an average offset as large as $10\%$ of the $J$ and $T_{max}$ is nearly unchanged.

\section{Experimental Implementations}
\label{impl} Spin models with the possibility of long time coherent dynamics are now being realized in some physical systems such as ultracold atoms in optical lattices and trapped ions. Here we outline the schematics of an implementation with ultracold atoms in optical lattices. Firstly a spin chain in an appropriate state has to be prepared. 1D tubes of light trapping atoms in a lattice are now quite routine \cite{impurity,bound}. The initial state can be prepared by dimerizing the lattice to a series of double wells. Simply dimerizing it would produce a series of Bell states as in Ref.\cite{Trotzky}, which would be the starting state for the scheme discussed in section \ref{BS}. Further, by having a magnetic field gradient between the two wells of the lattice, one can also prepare a series of $|\uparrow_k \downarrow_{k+1} \rangle$ or $|\downarrow_k \uparrow_{k+1} \rangle$ in double wells, and thereby the spin texture of one of the N\'eel states when many such wells are arranged in a series in a superlattice \cite{Trotzky}. The local control which is now possible in combination of digital spatial light modulators and microwave pulses \cite{impurity} should also enable one to prepare arbitrary spin textures if the results of section \ref{canted} are to be verified. The sudden quenching (switching on) of the interactions is done by lowering the barriers between neighboring wells at time $t=0$ in a time-scale much faster (say, a $ms$) than $\hbar/J$, which has been achieved quite recently \cite{impurity,bound}. In our setup, an open ended spin chain (in other words, a hard wall boundary) is required for the reflections. This should be achievable quite soon using the technique of digital light modulators which can eventually influence potentials at the scale of lattice site separations \cite{impurity}. The readout of spins of the end sites to verify the entanglement should be achievable using quantum gas microscope technique \cite{impurity}.
\section{Discussion and Summary}
Entangling distant spins is desirable for connecting separated quantum registers through teleportation. We find that it can be achieved by exploiting the dynamics of a spin chain after a quench. We find that obtaining a high entanglement can depend sensitively on the initial magnetic order. For example, an initial N\'eel state is an excellent resource, while more general canted order states hardly give any entanglement. Additionally, we found that a state composed of a product of singlets represent an excellent resource for generating entanglement when quenching the system by the $XX$ Hamiltonian. Our scheme can be implemented experimentally using, for example, ultracold atoms in optical lattices in which most of the requirements have been achieved lately, and an outline for the implementation is discussed in section \ref{impl}. Of course, the entanglement, while long range, is {\em not distance independent} in this scheme -- it falls with the length $N$ of the chain. This dependence, as well as the time needed to achieve the peak in entanglement, is plotted in Fig.\ref{Scaling}. Note that for an initial product of Bell states, for shorter chains, the entanglement produced is slightly lower than that produced from the N\'eel state -- though it catches up asymptotically as $N$ increases. Moreover, note from the slope of the line in Fig.\ref{Scaling}(b) that the time needed is always linear in $N$ i.e., $\sim \frac{N}{4J}$. With the velocity of a spin flip in the $XX$ chain being $\sim 2J$, this is the time needed for a spin flip to travel half the length of the chain. Note also from Fig.\ref{Scaling}(a) that even for very large chains ($N >100$) the maximally entangled fraction remains above $0.5$ (i.e., still useful for distillation).

Before concluding we provide the reader with some intuitive understanding of our results. Firstly, let us first clarify the non-triviality of our results. Basically, it is known (and expected) that a non-zero entanglement will develop between complementary blocks of a spin chain after a quench, simply because of entangled quasiparticles crossing the boundary between the blocks \cite{Calabrese,deChiara}. However, the spins at the two ends are ``non-complementary" parts of the chain. The part of the chain between the two spins serves as an environment. Thus for a state of the ends to be significantly entangled, it is required that the intervening chain be in the {\em same state} corresponding to distinct states of the end spins. To further clarify the situation, note that the Hamiltonian here conserves the number of spin flips. Thus states $|00\rangle$ and $|11\rangle$ of the end spins cannot have any coherence between them as they necessarily involve a different number of flips in the part of the chain between the spins. The only possibility of coherence is between the states $|01\rangle$ and $|10\rangle$ of the spin chain. However, though dynamics may produce the states $|01\rangle$ and $|10\rangle$ at the end spins, it is not guaranteed that the rest of the spin chain will be in the same state irrespective of whether the state $|01\rangle$ or $|10\rangle$ is generated between the end spins -- without that, there would be no entanglement between the end spins. It is highly nontrivial to ensure that the dynamics results in the central part of the chain being in the same state corresponding to $|01\rangle$ and $|10\rangle$ states. Because of reasons presented below, the Hamiltonian governing our dynamics and our initial state ensures this. It is additionally nontrivial that the amount of the entanglement between the end spins would be high. Why this is the case is explained in the following paragraphs.

We now proceed to explain the result that apart from states very close to the N\'eel state, any other canted order state gives a vanishing entanglement and fully entangled fraction. This is because of the principal mechanism through which entanglement is generated from the N\'eel state. For simplicity, we will present our explanation for odd $N$ chains. As shown in Fig.\ref{expl}, each up spin is the location of a fermion. The evolution is due to a free fermion model, so each fermion starts evolving in the chain as if other fermions were absent except for the phase factor the wave-function acquires upon pairwise exchange of fermions. Each fermion evolves in a superposition of left and right moving fermions, doing a so called quantum walk on the chain. This moving in a superposition of left and right creates, for example, from a configuration $|010\rangle$ in 3 successive sites a configuration of the form $|001\rangle+|100\rangle$, which, after factoring out the $|0\rangle$ of the central site, is an entangled state $|01\rangle+|10\rangle$. This is how each fermion acts as a source of an entangled state in the chain \citep{deChiara}. We will now justify why the simultaneous walk of all the fermions together then creates a highly entangled state between sites $1$ and $N$ at an optimal time. At time $t\sim \frac{N}{4J}$, where $2J$ is the velocity of fermions, the fermion at the exact centre of the chain will create, with some finite probability, a Bell state $|01\rangle+|10\rangle$ between the sites $1$ and $N$ because of the symmetry of the left and right walk. This is the $k=7$ fermion in Fig.\ref{expl}, which shows the distribution of this fermion after a quantum walk for a time $t\sim \frac{N}{4J}$. Note, from Fig.\ref{expl} that this fermion swaps its position with an equal number (in this case 3) of other fermions during its traversal to both the left and the right ends of the chain. Thus the relative phase between the left (i.e., $|10\rangle$) and the right (i.e., $|01\rangle$) components is $0$. The same holds for a fermion originating at any other site of the chain as both the left moving and the right moving components cross either an even number or an odd number of fermions (for example, in Fig.\ref{expl} the $k=3$ fermion crosses 1 fermion on the left and 5 fermions on the right to reach the ends). Moreover, the initial placement of fermions on odd sites of the chain also ensures that the relative phase between the single fermion transition amplitudes ($f_{k,1}$ and $f_{k,N}$) from their original site to the left and the right ends (when no other fermion is present) is also $0$ as each fermion acquires a factor of $e^{i\pi/2}$ per hop \footnote{This can be easily appreciated from the asymptotic form of the transition amplitudes $f_{kl} (t \sim \frac{N}{4J})\sim i^{k-l}J_{k-l}(t \sim \frac{N}{4J})$.}. For example, the $k=3$ fermion shown in Fig.\ref{expl} has 2 hops to reach the left end, giving a phase $e^{i\pi}$, while it has 10 hops to the right end, which gives a phase $e^{i 5 \pi}=e^{i \pi}$. Thus the fermions originating on an arbitrary site of the chain, would, in general, contribute entangled states of the form $\beta_1|01\rangle+\beta_N|10\rangle$ to the end sites (this contribution, in general, could be mixed with $|00\rangle$ and $|11\rangle$), where $\beta_1$ and $\beta_N$ can both be taken to be real and positive (the global phase outside $\beta_1|01\rangle+\beta_N|10\rangle$ does not matter as each fermion starts from a distinct location and has a distinct evolution, so that the entangled state stemming from each fermion is incoherently added up to those from the others to generate the final state of the two end spins). In general $\beta_1 \neq \beta_N$, as clear for the $k=3$ fermion in Fig.\ref{expl} (only for the central fermion, for example for $k=7$ in Fig.\ref{expl}, is $\beta_1=\beta_N$). Thus the coefficient of the off-diagonal term $|01\rangle \langle 10|$ contributed by each fermion is $\beta^{*}_1 \beta_N$, which is positive, and add up together to make a substantial off-diagonal term of the total density matrix. The entanglement is substantial because this off-diagonal term is substantial. Now imagine that there is a small probability of $\epsilon$ for a fermion to be absent from a site were it was supposed to be present. This is a small deviation from the N\'eel state. This causes a state of the form $\beta_1|01\rangle-\beta_N|10\rangle$ with $\beta_1$ and $\beta_N$ both real and positive, to be generated because if the left moving component of a fermion's state crosses an even number of other fermions then the right moving component crosses an odd number of other fermions and vice-versa. In Fig.\ref{expl}, for example, if the fermion at the $k=5$ site was absent, then the central $(k=7)$ fermion would cross 2 fermions on its left, while it would still cross 3 fermions to its right. This will reduce the off-diagonal term by subtracting a proportion $\epsilon \beta_1^*\beta_N$ from it. A very similar result holds when an extra fermion is present at a site were it was not supposed to be. Imagine, for example, the state of the $k=4$ site in Fig.\ref{expl} was up instead of being as depicted. While this extra fermion itself will still create a state $\beta_1|01\rangle+\beta_N|10\rangle$, as can be verified by counting its number of hops and fermion crossings in reaching the ends, {\em all} other fermions will now contribute a state
$\beta_1|01\rangle-\beta_N|10\rangle$, by virtue of one extra fermion being now being crossed on either its leftward or its rightward walk to the ends. Thus overwhelmingly the presence of an extra fermion still results in contributions $-\epsilon \beta^{*}_1 \beta_N$ to the original state. As there are $N$ sites in which a single spin flip error can occur, as soon as $\epsilon \sim \frac{1}{N}$, the contributions of the form $-\beta_1^{*}\beta_N$ become of the order of unity and cancel contributions of the form $\beta^{*}_1\beta_N$ arising from the error-free N\'eel state. Thus for very low errors of the magnitude $\epsilon \sim \frac{1}{N}$ about the N\'eel state, the entanglement completely vanishes. The canted ordered state, in fact, can be regarded as quite a systematic error whose probability $\epsilon$ increases from the left to the right of the chain, and thereby $\Delta \alpha$ falls so rapidly with $N$. Thus the generated entanglement is quite rapidly lost as canting angles start deviating from the N\'eel ordered state. The starting resource of Bell states is a case were we are starting with states $|01\rangle-|10\rangle$ on neighboring sites, which is as entangled as the state that is created after a one step (left and right in superposition) evolution of a single fermion from the N\'eel state. The similarities of the resources near the starting point of the evolution is thereby responsible for both the product of Bell states and the N\'eel state being excellent resources for establishing entanglement.

\section{Acknowledgements} BA is supported by King Saud University. HW was supported by a UCL PhD plus fellowship when this work was started. The work of SB is supported by the ERC grant PACOMANEDIA. We wish to thank Abolfazl Bayat and Marcus Cramer for discussions that form a background to this work.

\end{document}